\documentclass[12pt]{article}

\usepackage{ulem}
\usepackage{comment}

\usepackage[dvips]{color,graphicx}
\DeclareGraphicsExtensions{.pdf,.jpeg,.png,.jpg,.eps}
\usepackage[top=0.95in, bottom=0.95in, left=0.95in, right=0.95in]{geometry}
\usepackage{subfig}
\usepackage{amssymb}
\usepackage{amsfonts}
\usepackage{amsmath}
\usepackage{setspace}
\usepackage{epstopdf}
\usepackage{mathtools}
\usepackage{pstricks}
\usepackage{bbold}
\usepackage{authblk}
\usepackage{bm,bbm}
\usepackage{utopia}

\usepackage{theorem}
\usepackage{lscape}
\usepackage{xr-hyper}
\usepackage[colorlinks, linkcolor=blue, anchorcolor=blue, citecolor=blue, breaklinks]{hyperref}

\usepackage {natbib}
\newcommand{\bV}{\mbox{\boldmath $b$}}

\newcommand{\eV}{\mbox{\boldmath $e$}}
\newcommand{\fV}{\mbox{\boldmath $f$}}

\newcommand{\sV}{\mbox{\boldmath $s$}}

\newcommand{\mV}{\mbox{\boldmath $m$}}
\newcommand{\xV}{\mbox{\boldmath $x$}}
\newcommand{\uV}{\mbox{\boldmath $u$}}
\newcommand{\yV}{\mbox{\boldmath $y$}}
\newcommand{\zV}{\mbox{\boldmath $z$}}
\newcommand{\QV}{\mbox{\boldmath $Q$}}

\newcommand{\YV}{\mbox{\boldmath $Y$}} 

\newcommand{\ZV}{\mbox{\boldmath $Z$}}

\newcommand{\deltaB}{\mbox{\boldmath $\delta$}}

\newcommand{\betaB}{\mbox{\boldmath $\beta$}}

\newcommand{\lambdaB}{\mbox{\boldmath $\lambda$}}
\newcommand{\gammaB}{\mbox{\boldmath $\gamma$}}

\newcommand{\rhoB}{\mbox{\boldmath $\rho$}}

\newcommand{\phiB}{\mbox{\boldmath $\phi$}}

\newcommand{\sumiN}{\sum_{i=1}^N}

\newcommand{\fracNT}{\frac1{NT}}
\newcommand{\tr}{\mathrm{tr}}

\newcommand{\RNum}[1]{\uppercase\expandafter{\romannumeral #1\relax}}

\newcommand{\calP}{\mathcal P}

\newcommand{\calX}{\mathcal X}

\newcommand{\bmQ}{\bm{Q}}

\newcommand{\ol}[1]{\overline{#1}}

\newtheorem{thm}{Theorem}

\newtheorem{remark}{Remark}
\newtheorem{proposition}{Proposition}
\allowdisplaybreaks[4]

\title{Bayesian inference for dynamic spatial quantile models with interactive effects
\thanks{
The paper has benefited from constructive comments from participants at Australian spatial econometrics and statistics workshop 2023, held at Monash university.
This research was supported by University of Melbourne Research Computing Services and the Petascale Campus Initiative.
It is also supported by the Australian Research Council Discovery Grants DP230100959 and DP240101009.
}}

\author[1]{Tomohiro Ando
\thanks{
A part of this work was completed when TA was visiting Monash University.
TA would like to thank their generous supports during the visit.
}}
\author[2]{Jushan Bai}
\author[3]{Kunpeng Li}
\author[4]{Yong Song}
\affil[1]{The University of Melbourne Melbourne Business School}
\affil[2]{Columbia University}
\affil[3]{Capital University of Economics and Business}
\affil[4]{University of Melbourne}

\begin{document}

\maketitle

\begin{abstract}
\baselineskip=15pt

This paper proposes a dynamic spatial panel quantile model that accounts for unobserved heterogeneity. The Bayesian MCMC algorithm developed here introduces several novel features tailored to the challenges of high-dimensional dynamic spatial quantile models. Key innovations include quantile randomisation, which enables conditional conjugacy and facilitates efficient sampling, and a new Gibbs sampler specifically designed to handle structural spatial parameters. We also establish the Bayesian consistency of the proposed estimation method. Finally, we demonstrate the practical utility of the approach through an application to the quantile co-movement structure of the Australian gasoline market.

\end{abstract}

\vspace{4mm}



\newpage

\section{Introduction}

\baselineskip=23pt

Spatial panel data models explore spatial interactions across individuals or regions over time. These models have found wide-ranging applications in fields such as housing economics (e.g., \cite{BF2015}), marketing (e.g., \cite{hunneman2022}), and urban economics (e.g., \cite{Glaser2022}). In this paper, we introduce a novel dynamic spatial panel quantile model with interactive effects. This model simultaneously accounts for spatial dependence, temporal dynamics, unobserved heterogeneity, and heterogeneous effects across quantiles—features that are often observed in economic and social data but rarely modeled together.

There is a rich literature on linear panel models with interactive effects (e.g., \cite{bai2009, baili2014}; \cite{ando2017}; \cite{hallin2007}; \cite{moon2015}; \cite{pesaran2006}; \cite{stock2002}; \cite{lu2016}) and linear spatial panel models (e.g., \cite{aquaro2021}; \cite{baltagi2011}; \cite{bai2015}; \cite{kelejian2004}; \cite{lee2004}; \cite{lin2010}; \cite{li2017}; \cite{lu2017}; \cite{qu2015}; \cite{reich2011}; \cite{shi2017}; \cite{yu2008}). However, research on quantile regression (see \cite{KB1978}) in spatial and high-dimensional panel settings remains limited, especially for models that incorporate both interactive effects and dynamic structures.
 Recent work by \cite{ando2020} extended quantile methods to panel data with interactive effects, capturing quantile co-movements while allowing for heterogeneous slope coefficients. However, their model does not accommodate dynamic and spatial dependencies in quantiles.


Spatial interactions and temporal dynamics are fundamental features of economics and the social sciences. Spatial dependence may arise from peer effects, spillover effects, or imitative behavior (see \cite{manski1993, anselin1988}, among others), while temporal dependence often reflects persistence in individual or group behavior over time. To capture these complex features, we propose a dynamic spatial quantile panel model with interactive effects. This framework facilitates quantile-specific analysis of both spatial and temporal dependence, accommodates heterogeneous slope coefficients, and incorporates latent factor structures. Due to the model’s high-dimensional parameter space—driven by the dynamic structure and large number of cross-sectional units—we develop a new Bayesian Markov Chain Monte Carlo (MCMC) estimation method tailored to this setting.

{\bf Methodological challenges}: Several methodological challenges arise in the estimation process. First, because the quantile function is defined implicitly in terms of its coefficients, a conventional Gibbs sampler is not directly applicable. To overcome this, we introduce randomised quantile dynamics that create conditional conjugacy, allowing for efficient sampling. Second, the spatial autoregressive parameter is structural in nature and does not permit a standard Gibbs step. Inspired by methods in structural vector autoregression, we construct a bimodal mixture distribution to incorporate this parameter within our MCMC framework. Finally, to alleviate the computational burden from high-dimensional matrix operations, we apply a breadth-first search algorithm from graph theory to block-diagonalise the spatial matrix, thereby significantly reducing computational complexity. These methodological innovations enable our MCMC algorithm to efficiently estimate large-scale dynamic spatial quantile panel models with latent factor structures.

{\bf Theoretical Challenges:}
To establish the validity of our Bayesian approach, we prove Bayesian consistency—a theoretical result not previously achieved for panel data models with interactive effects. This requires overcoming several substantial hurdles: the dynamic structure of the model, high-dimensional incidental parameters from latent factors and their loadings, the nonsmooth nature of the quantile objective function, the nonlinearity introduced by the spatial term, and the rotational indeterminacy of the factor structure. Although recent work (e.g., \cite{ando2020}; \cite{chen2020}) addresses some related issues, their results are not applicable to our setting. Consequently, we first derive new asymptotic properties within a frequentist framework (see Remark 7), which we then build upon to develop our Bayesian theory. Even with these foundations, our model demands additional theoretical innovation to rigorously establish Bayesian consistency.

Our contributions are summarised as follows. 
First, we introduce a dynamic spatial panel quantile model with interactive effects, designed to accommodate key features such as spatial dependence and temporal dynamics.
This specification enhances the model’s flexibility in capturing complex data patterns. Second, we develop a novel Bayesian estimation procedure tailored to high-dimensional parameter spaces. Third, we establish several asymptotic properties, including Bayesian consistency. Finally, we demonstrate the utility of our framework through an empirical analysis of the Australian gasoline market.

The paper is organised as follows.
Section 2 introduces a new dynamic spatial panel quantile model with interactive fixed effects.
Section 3 outlines the proposed Bayesian MCMC estimation procedure.
Section 4 establishes the Bayesian consistency of the method.
In Section 5, we apply the proposed approach to the Australian gasoline market.
Section 6 concludes the paper.
To conserve space, all technical proofs are provided in the online supplementary material. The supplementary document also includes Monte Carlo simulation results, which demonstrate the effectiveness of the proposed estimation procedure.

{\bf Notations:}
Let $\| A\|$ = $[\tr(A'A)]^{1/2}$ be the Frobenius norm of matrix $A$, where ``tr'' denotes the trace of a square matrix, and let $\|A\|_2$ be its spectrum norm (the largest singular value of $A$).
In addition, for any $N\times N$ matrix
$\|A\|_1$ is defined as $\|A\|_1=\max_{1\le j \le N}$ $\sum_{i=1}^N |a_{ij}|$
where $a_{ij}$ is the $(i, j)$-th element of $A$.
Similarly,
$\|A\|_\infty=\max_{1\le i \le N}\sum_{j=1}^N |a_{ij}|$.
For sequences $a_n$ and $b_n$, the notation $a_n \lesssim b_n$ means $a_n = O(b_n)$, that is, there exists $C>0$ and for all $n$ large enough, $a_n \le C b_n$.
We write $c_n = O_p(d_n)$ if  $c_n/d_n$ is stochastically bounded, and $c_n = o_p(d_n)$ if $c_n/d_n$ converges to zero in probability.

\section{Dynamic spatial quantile models with
interactive effects}

Suppose that, for the $i$-th unit $(i=1,...,N)$ at time $t$ $(t=1,...,T)$, its response $y_{it}$ is observed together with a set of
$p$ explanatory variables $\{x_{it,1},...,x_{it,p}\}$.
We consider the $\tau$-th quantile function of $y_{it}$ by jointly modeling spatial effects, time effects and common shocks.
To capture these effects simultaneously, we define the $\tau$-th quantile function of $y_{it}$ as
\begin{align}
&Q_{y_{it}}
\Big(\tau|X_t, F_{t,\tau},B_\tau,\Lambda_\tau,\rhoB_\tau,\gammaB_\tau,\deltaB_\tau\Big)
\nonumber\\
&\equiv
\rho_{i,\tau}
\sum_{i\neq j,j=1}^Nw_{ij}
Q_{y_{jt}}\Big(\tau|X_t, F_{t,\tau},B_\tau,\Lambda_\tau,\rhoB_\tau,\gammaB_\tau,\deltaB_\tau\Big) \notag\\
&\quad +
\delta_{i,\tau}
\sum_{i\neq j,j=1}^Nw_{ij}
Q_{y_{j,t-1}}\Big(\tau|X_{t-1}, F_{t-1,\tau},B_\tau,\Lambda_\tau,\rhoB_\tau,\gammaB_\tau,\deltaB_\tau\Big)
\nonumber\\
&\quad+
\gamma_{i,\tau}
Q_{y_{i,t-1}}\Big(\tau|X_{t-1}, F_{t-1,\tau},B_\tau,\Lambda_\tau,\rhoB_\tau,\gammaB_\tau,\deltaB_\tau\Big)
+ \xV_{it}'\bV_{i,\tau}+\fV_{t,\tau}'\lambdaB_{i,\tau}
\label{model}
\end{align}
for $i=1, \dots ,N$ and $t=1, \dots ,T$.
Here
$w_{ij}$ $(i=1,2,\cdots,N; j=1,2,\cdots,N)$ are pre-specified spatial weights with $w_{ii}=0$,
$\rho_{i,\tau}$ and $\delta_{i,\tau}$ are the heterogeneous spatial parameters capturing the strength of the spillover effects,
the coefficients $\gamma_{i,\tau}$ are the heterogeneous temporal parameters,
$\xV_{it}=(1,x_{it,1},...,x_{it,p})'$ is $(p+1)$ -dimensional vector of explanatory variables;
$B_{\tau}=(\bV_{1,\tau}, \bV_{2,\tau}, \dots, \bV_{N,\tau})'$,
$\bV_{i,\tau}=(b_{i,0,\tau},b_{i,1,\tau},...,b_{i,p,\tau})'
$
is
a $(p+1)$-dimensional vector of regression coefficients;
$\fV_{t,\tau}=(f_{t1,\tau},...,f_{t r_\tau,\tau})'$ is $r_{\tau}$-dimensional unobservable common factors;
$\lambdaB_{i,\tau}=(\lambda_{i1,\tau},...,\lambda_{i r_\tau,\tau})'$ is $r_{\tau}$-dimensional vector of factor loadings;
$X_t$ and $F_{t,\tau}$ are information on the explanatory variables and the common factors up to time $t$.
The exact expression of $\xV_{it}'\bV_{i,\tau}$ is
\begin{equation}\label{eq:eit}
\xV_{it}'\bV_{i,\tau}
=
\sum_{k=1}^{p} x_{it,k}b_{ik,\tau}
+G_{i,e_{it}}^{-1}(\tau),
\end{equation}
where
$e_{it}$ is the idiosyncratic error term and $G_{i,e_{it}}^{-1}(\tau)$ is the $\tau$-th quantile point of $e_{it}$ with $G_{i,e_{it}}(\cdot)$ being the cumulative distribution function of $e_{it}$.
Thus, the $\tau$-th quantile of the idiosyncratic error $G_{i,e_{it}}^{-1}(\tau)$, which depends only on $i$ and $\tau$, is absorbed by the term $\xV_{it}'\bV_{i,\tau}$ since the first element of $\xV_{it}$ is 1.
We assume that $e_{it}$ is identically distributed over $t$ while its distribution may vary over $i$.

The preceding quantile function (\ref{model}) is associated with the following data-generating process, provided that the right-hand side of the equation is an increasing function of $u_{it}$,
\begin{eqnarray*}
y_{it, u_{it}}
=
\rho_{i,u_{it}}
\sum_{j=1}^Nw_{ij} y_{jt, {u}_{it}}
+
\gamma_{i,{u}_{it}} y_{i,t-1,{u}_{it}}
+
\delta_{i,{u}_{it}}
\sum_{j=1}^Nw_{ij} y_{j,t-1, {u}_{it}}
+\xV_{it}'\bV_{i,{u}_{it}}+\fV_{t,{u}_{it}}'\lambdaB_{i,{u}_{it}},
\end{eqnarray*}
where
$y_{it, u_{it}}$ is $\tau={u}_{it}$-th quantile and ${u}_{it}$  are i.i.d. $U(0,1)$. The coefficient of the constant regressor absorbs the error term. As seen, we integrate spatial interactions and temporal dynamics into a model, which allows for both contemporaneous and dynamic spatial effects, thereby enabling the capture of temporal spillover effects and peer influences in the spatial domain.

{\remark
\cite{koenker2006} considered autoregressive quantile model in the context of univariate time series. Their quantile function is expressed as a weighted sum of past observed values of the response variable. While their model is regarded as autoregressive in this sense,   the quantile function itself is not autoregressive. In contrast, our quantile function in (\ref{model}) includes a weighted sum of past quantile functions, making the quantile itself autoregressive, or dynamic.
}

\bigskip

Define the $N\times N$ matrix $S(\rhoB_{\tau})\equiv (I-\rhoB_{\tau} W)^{-1}$, where
$\rhoB_{\tau}=\mathrm{diag}(\rho_{1,\tau}, \dots, \rho_{N,\tau})$, and
$W=[w_{ij}]$ is the $N\times N$ spatial weights matrix.
Also, we define
the $N\times N$ matrix $A(\rhoB_{\tau},\deltaB_{\tau},\gammaB_{\tau})\equiv (I-\rhoB_{\tau} W)^{-1}(\gammaB_{\tau}+\deltaB_{\tau} W)$ where
$\deltaB_{\tau}=\mathrm{diag}(\delta_{1,\tau}, \dots, \delta_{N,\tau})$,
$\gammaB_{\tau}=\mathrm{diag}(\gamma_{1,\tau}, \dots, \gamma_{N,\tau})$.
Stack the quantile functions over cross sections by defining
\[\bm{Q}_t\Big(X_t, F_{t,\tau},B_\tau,\Lambda_\tau,\rhoB_\tau,\gammaB_\tau,\deltaB_\tau\Big)=\begin{bmatrix}
Q_{y_{1t}}\big(\tau| X_t, F_{t,\tau},B_\tau,\Lambda_\tau,\rhoB_\tau,\gammaB_\tau,\deltaB_\tau\big)\\
\vdots \\
Q_{y_{Nt}}\big(\tau| X_t, F_{t,\tau},B_\tau,\Lambda_\tau,\rhoB_\tau,\gammaB_\tau,\deltaB_\tau\big)
\end{bmatrix}\]
By recursive substitution, model (\ref{model}) can be rewritten as
\begin{align*}
\bm{Q}_t\Big(X_t, F_{t,\tau},B_\tau,\Lambda_\tau,\rhoB_\tau,\gammaB_\tau,\deltaB_\tau\Big)
&=
\sum_{h=0}^{t-1}
P_h(\rhoB_{\tau},\deltaB_{\tau},\gammaB_{\tau})
\begin{bmatrix}
\xV_{1,t-h}'\bV_{1,\tau}+\fV_{t-h,\tau}'\lambdaB_{1,\tau}
\\
\vdots \\
\xV_{N,t-h}'\bV_{N,\tau}+\fV_{t-h,\tau}'\lambdaB_{N,\tau}
\end{bmatrix},
\nonumber
\end{align*}
where
the $N\times N$ matrix $P_h(\rhoB_{\tau},\deltaB_{\tau},\gammaB_{\tau})$
is given as
\begin{eqnarray}
P_h(\rhoB_{\tau},\deltaB_{\tau},\gammaB_{\tau})\equiv
A(\rhoB_{\tau},\deltaB_{\tau},\gammaB_{\tau})^h
S(\rhoB_{\tau}).
\label{Pmat}
\end{eqnarray}
We will impose restrictions on (\ref{Pmat}) to ensure the sum  to be well defined (referred to as stationarity).  In addition, we assume $\bm{Q}_t(\cdot)=0$ when $t=0$. The effect of the initial condition is generally negligible if $T$ is large.
Thus, an alternative expression of (\ref{model}) is
\begin{eqnarray}
Q_{y_{jt}}\Big(\tau| X_t,B_{\tau}, F_{t,\tau}, \Lambda_{\tau}, \rhoB_{\tau},\gammaB_\tau,\deltaB_\tau\Big)=
\sum_{h=0}^{t-1}\sum_{k=1}^N
p_{jk,h}(\rhoB_{\tau},\deltaB_{\tau},\gammaB_{\tau})\Big(\xV_{k,t-h}'\bV_{k,\tau}+\fV_{t-h,\tau}'\lambdaB_{k,\tau}\Big),
\label{modelreduced}
\end{eqnarray}
where  $p_{ij,h}(\rhoB_{\tau},\deltaB_{\tau},\gammaB_{\tau})$ is the $(i,j)$th element of $P_h(\rhoB_{\tau},\deltaB_{\tau},\gammaB_{\tau})$ in (\ref{Pmat}).

To eliminate the rotational indeterminacy of the common factor structure, we need to impose a restriction on $F_\tau=(\fV_{1,\tau},...,\fV_{T,\tau})'$ and $\Lambda_\tau=(\lambdaB_{1,\tau},...,\lambdaB_{N,\tau})'$.
For example, \cite{bai2013} imposed the followings
\begin{eqnarray}
\frac1TF_\tau'F_\tau=I_{r_\tau}~~~{\rm and}~~~\frac1N\Lambda_\tau'\Lambda_\tau=D_{r_{\tau}},
\label{ID}
\end{eqnarray}
where $I_{r_{\tau}}$ is an $r_{\tau}\times r_{\tau}$ identity matrix, and $D_{r_{\tau}}$ is a diagonal matrix whose diagonal elements are distinct and are arranged in a descending order.
We refer to \cite{baing2013} for alternative restrictions on the common factor structure.

We have to estimate the unknown parameters
$\rhoB_{\tau}$,
$\gammaB_\tau$,
$\deltaB_\tau$,
$B_\tau$,
$\Lambda_\tau$, and $F_\tau$ simultaneously.
Let $\vartheta_\tau=\{\rhoB_\tau, \deltaB_\tau, \gammaB_\tau, B_\tau, \Lambda_\tau, F_\tau\}$.
The frequentist estimator can be obtained as the minimiser of the following objective function
\begin{eqnarray}
\ell_\tau(Y|X,\vartheta_\tau)
=\frac{1}{NT}\sum_{i=1}^N\sum_{t=1}^Tq_\tau
\left(y_{it}-Q_{y_{it}}\left(\tau|X_t, \vartheta_\tau\right)\right)
\label{Allloss}
\end{eqnarray}
where
$Q_{y_{it}}\left(\tau|X_t, \vartheta_\tau\right)\equiv
Q_{y_{it}}(\tau|X_t, F_{t,\tau},B_\tau,\Lambda_\tau,\rhoB_\tau,\gammaB_\tau,\deltaB_\tau)$ is defined in (\ref{model}),
$q_\tau (u)=u(\tau-I(u\le 0))$ is the quantile loss function, $Y \equiv \{y_{it}|i=1,...,N,t=1,...,T\}$ and $X \equiv \{\xV_{it}|i=1,...,N,t=1,...,T\}$.

\section{Bayesian estimation}

In this paper, we employ Bayesian MCMC sampling procedure to produce posterior samples from the posterior distribution
\begin{eqnarray}
\pi(\vartheta_\tau|X,Y)
\propto
\exp\left(-\ell_\tau(Y|X,\vartheta_\tau)
\right)
\pi(\vartheta_\tau),
\label{Posterior}
\end{eqnarray}
where $\pi(\vartheta_\tau)$ is a prior density of the parameter $\vartheta_\tau$.
Details of $\pi(\vartheta_\tau)$ are given below.

The quantile structure in (\ref{modelreduced}) is
a highly nonlinear function of parameters $\rhoB_\tau, \gammaB_\tau$, $\deltaB_\tau$ and $B_\tau$.
Such nonlinearity costs the conditional conjugacy for inference,
rendering the ideal Gibbs sampling infeasible.
Simple methods such as the Metropolis-Hastings might
be very inefficient due to the large number of parameters.
For example, in our application,
the dimension of the parameter space is greater than 44,000.
In this paper, we propose to randomise the quantile equation
to adapt to the Bayesian inference framework.
In particular,  we set
\begin{align}
\varepsilon_{it,\tau} &=y_{it}-Q_{y_{it}}(\tau| X_t,\vartheta_\tau), \label{eq:sq:y}\\
Q_{it,\tau}& = \rho_{i,\tau} \sum\limits_{j=1, j\neq i}^N w_{ij} Q_{jt,\tau}
 + \gamma_{i,\tau} Q_{i,t-1,\tau} +\delta_{i,\tau} \sum\limits_{j=1, j\neq i}^N w_{ij} Q_{j,t-1,\tau}+ \xV_{it}' \bV_{i,\tau} + \fV_{t,\tau}'\lambdaB_{i,\tau}  + e^q_{it,\tau}.\label{eq:sq:q}
\end{align}
In the Bayesian framework, it is commonly assumed that $\varepsilon_{it,\tau}$
in \eqref{eq:sq:y}
 has an Asymmetric Laplace distribution (ALD):
$p(\varepsilon_{it,\tau}\mid\tau,\sigma )
=\frac{\tau(1-\tau)}{\sigma}\exp(-\frac{q_\tau(\varepsilon_{it,\tau})}{\sigma})
$.
The Gibbs sampling method of \cite{kozumi2011gibbs}uses a scaled mixture representation of the ALD. Under this representation,
the conditional density of $y_{it}$, given the auxiliary variable $V_{it}$,
is expressed as normal:
\begin{eqnarray}
p(y_{it} \mid X_t, \vartheta_\tau,V_{it}, \sigma )
\propto \sqrt{\frac{\tau(1-\tau)}{2\sigma^2V_{it}}}
\exp\left(-\frac{(y_{it}-Q_{y_{it}}(\tau| X_t,\vartheta_\tau)-\frac{1-2\tau}{\tau(1-\tau)}\sigma V_{it})^2}{\frac{4}{\tau(1-\tau)}\sigma^2V_{it}}\right),
\label{ALD}
\end{eqnarray}
where
$V_{it}\sim Exp(1)$ and
$\sigma$ is the temperature parameter
to facilitate Bayesian inference.



We denote $Q_{y_{it}}(\tau|X_t, \vartheta_\tau)$ as $Q_{it,\tau}$ in (\ref{eq:sq:q}), and
$e^q_{it,\tau}\sim N(0, \sigma_{q,\tau}^2)$.
Note that $e_{it,\tau}$ in (\ref{eq:eit}) and $e_{it,\tau}^q$ in (\ref{eq:sq:q}) are different.
The introduction of $e^q_{it,\tau}$
is the randomisation that allows to treat the quantile values as a latent variable
to augment the parameter space.
 Without $e^q_{it,\tau}$,
the $Q_{it, \tau}$ is a deterministic function of $\vartheta_\tau$
in \eqref{eq:sq:q}
and hence cannot be treated as conditioning parameters.
Consequently, the pseudo posterior
kernel \eqref{Posterior} will be a complicated function of $\vartheta_\tau$,
making posterior computation burdensome, particularly in high-dimensional settings.
By introducing the randomised component $e^q_{it,\tau}$, the process  $Q_{it,\tau}$ gains variability and can be considered part of the augmented parameter space. This randomisation enables efficient Gibbs sampling by allowing the conditional distributions to take tractable forms.
In this way, equation~\eqref{eq:sq:q} plays a key role in making Gibbs sampling computationally feasible. See also our discussion in Remark~3.



\subsection{Prior setting}

We apply the same prior to the parameters for each quantile $\tau$.
So ignore $\tau$ for notational simplicity.
The parameter space includes:
\begin{enumerate}
\item A normal prior is used for the lag parameter $\gamma_i\sim N(m_\gamma, h_\gamma^{-1}),~i=1,2,...,N$.

\item A normal prior is used for the spatial lag parameter $\delta_i\sim N(m_\delta, h_\delta^{-1}),~i=1,2,...,N$.

\item A multivariate normal prior is used for $\bV_i\sim N(\mV_b, H_b^{-1}),~i=1,2,...,N.$

\item A normal prior is used for the spatial parameter $\rho_i \sim N(m_\rho,  h_{\rho}^{-1}),~i=1,2,...,N$.

\item For the common factor, we assume a stationary AR(1) process as
$f_{j,t}=\phi_j f_{j,t-1}+e^f_{j,t}$,
for $j=1,...,r$, where $e^f_{j,t}\sim N(0, 1)$ with a unit variance for identification purpose.
Each $e^f_{j,t}$ is independent over $j$ and $t$.
Also, we assume the initial condition
$f_{j, 1}\sim N(0, h_f^{-1}),$
for $j=1,...,r$.

\item For the autoregressive coefficient, we set
$\phi_j\sim U(-1, 1),~j=1,...,r$.

\item
For the factor loading $\lambda_{ij}$, we use a  normal prior:
$\lambda_{ij}\sim N(0, h_\lambda^{-1})$.
The first $ r\times r$ block is a lower triangular matrix such that
$\lambda_{ii}\sim N(0, h_\lambda^{-1}) \textbf{1}(\lambda_{ii}>0)$
for $i=1,...,r$, and  $\lambda_{ij}=0$ if $j>i$.

\item Gamma prior
$\sigma\sim G(v_\sigma, s_\sigma)$ is used for
$\sigma$ in the asymmetric Laplace error term.


\end{enumerate}


{\remark

In addition to $\vartheta_\tau$ in the model (\ref{model}),
the Bayesian parameter space also has auxiliary variables
$\vartheta_\tau'=\{Q_\tau, V, \phiB, \sigma\}$,
where $V$ and $Q_\tau$ are the collections of $V_{it}$ in (\ref{ALD}) and $Q_{it,\tau}$, respectively.
Denote the corresponding posterior as
\begin{eqnarray}
p(\vartheta_\tau, \vartheta_\tau'\mid X, Y)
\propto
p(Y|X,\vartheta_\tau,V,\sigma)
\pi(\vartheta_\tau)
\pi(\vartheta_\tau')
\label{PseudoPosterior}
\end{eqnarray}
where
$p(Y\mid X, \vartheta_\tau, V,\sigma)=\prod_{i=1}^N\prod_{t=1}^T
p(y_{it} \mid X_t, \vartheta_\tau,V_{it} , \sigma)
$ with $p(y_{it} \mid X_t, \vartheta_\tau,V_{it} , \sigma)
$  given in (\ref{ALD}), and
$\pi(\vartheta_\tau')$ is a prior density of $\vartheta_\tau'$.
The concentrated posterior for comparison with the frequentist parameter set
is defined as $p(\vartheta\mid X, Y) = \int p(\vartheta, \vartheta'\mid X, Y)d\vartheta'$.
Although $p(\vartheta\mid X, Y)$ here and $\pi(\vartheta\mid X, Y)$ in (\ref{Posterior}) are different, Proposition 1 in Section 3.2 below  ensures that our MCMC posterior samples generated from $p(\vartheta\mid X, Y)$ can be regarded as those generated from $\pi(\vartheta\mid X, Y)$ through the importance sampling procedure.

}

\subsection{MCMC sampling procedure}
\label{sec:mcmc}

We briefly describe the procedure in this section,
with detailed techniques available in the Appendix.
The computational challenge of our model arises from three main factors. First, the high dimensionality
of the quantile values and their associated dynamics require a significant number of large matrix inversions.
Second, due to the substantial heterogeneity in the model, generic methods, such as Metropolis-Hastings or particle filters, face high computational cost.
Finally, the large dimension \( N \) and time period \( T \) result in a
vast number of observations, further intensifying the computational burden. Consequently, our methods are designed to rely on the Gibbs sampler whenever possible.

The Markov Chain Monte Carlo (MCMC) algorithm for posterior inference proceeds as follows.
Each step is performed conditionally on all other parameters. For clarity,
subscripts are omitted, and we present the key elements below; full details are provided in the supplementary material.

\begin{enumerate}
    \item $\gamma_i \sim N(\ol{m}_{\gamma_i}, \ol{h}^{-1}_{\gamma_i})$,
    where $\ol{h}_{\gamma_i}=h_{\gamma}+\sigma^{-2}_q(\xV^*_i)'\xV_i^*$ and
    $\ol{m}_{\gamma_i} = \ol{h}^{-1}_{\gamma_i}(h_\gamma m_\gamma + \sigma^{-2}_q(\xV_i^*)'\yV_i^* )$,
    $\yV_i^*=(y^*_{i1},...,y^*_{it})'$, $\xV_i^*=(x^*_{i1},...,x^*_{iT})'$,
    $y^*_{it}=Q_{it}- [\rho_{i} \sum_{j=1, j\neq i}^N w_{ij} Q_{jt}+
    \delta_{i} \sum_{j=1, j\neq i}^N w_{ij} Q_{j,t-1} + \xV_{it}' \bV_{i} + \fV_{t}'\lambdaB_{i}  ]$
    and  $x^*_{it} = Q_{i,t-1}  $.

    \item $\delta_i\sim N(\ol{m}_{\delta_i}, \ol{h}^{-1}_{\delta_i})$,
    where $\ol{h}_{\delta_i}=h_\delta+\sigma^{-2}_q(\xV^*_i)'\xV_i^*$ and $\ol{m}_{\delta_i} = \ol{h}_{\delta_i}^{-1}(h_\delta m_\delta + \sigma^{-2}_q(\xV_i^*)'\yV_i^* )$,
    $\yV_i^*$ and $\xV_i^*$ are the stacking of $y^*_{it}$ and $x^*_{it}$,
    respectively, with $y^*_{it}=Q_{it}- [\rho_{i} \sum_{j=1, j\neq i}^N w_{ij} Q_{jt}+\gamma_{i} Q_{i,t-1} + \xV_{it}' \bV_{i} + \fV_{t}'\lambdaB_{i}  ]$
    and  $x^*_{it} = \sum_{j=1, j\neq i}^N w_{ij} Q_{j,t-1}  $.

    \item $\bV_i \sim N(\ol{\mV}_{b_i}, \ol{H}^{-1}_{b_i}),$
    where $\ol{H}_{b_i}=H_b+\sigma^{-2}_q(X^*_i)'X_i^*$ and $\ol{\mV}_{b_i} = \ol{H}_{b_i}^{-1}(H_b \mV_b + \sigma^{-2}_q(X_i^*)'\yV_i^* )$.
    The vectors $\yV_i^*$ and matrix $X_i^*$ are the stacking of $y^*_{it}$ and $x'_{it}$,
    respectively with
    $y^*_{it}=Q_{it}- [\rho_{i} \sum_{j=1, j\neq i}^N w_{ij} Q_{jt}+ \gamma_{i} Q_{i,t-1} +\delta_{i} \sum_{j=1, j\neq i}^N w_{ij} Q_{j,t-1} + \fV_{t, \tau}'\lambdaB_{i,\tau}  ]$ while $x_{it}$ is the data.

    \item $\rho_i$ does not have a conjugate representation, $\rho_i$ is a structural parameter from a high-dimensional simultaneous equations system.
    We propose to draw $\rho_i$
    from a two-component mixture normal distribution
    $\rho_{i}\sim w_i N(\mu_{1i},\sigma^2_{1i})+(1-w_i)N(\mu_{2i}, \sigma^2_{2i})$
    to approximate the posterior $p(\rho_i\mid \cdot)$,
    as being inspired by \cite{villani2009steady}.
    The mixture weight $w_i$ and parameters $\mu_{1i}, \mu_{2i},\sigma^2_{1i},\sigma^2_{2i}$
    are detailed in the Supplement A1.

    \item The factor $\fV_t$ has a state space representation.
    The measurement equation is
    $\yV^*_t=  \Lambda \fV_t + \eV^q_t$ with $ \eV^q_t\sim N(0, \sigma^2_q I_N)$,
    where $\yV^*_t=(y^*_{1t},...,y^*_{Nt})$
    with  $y^*_{it}=Q_{it}- [\rho_{i} \sum_{j=1, j\neq i}^N w_{ij} Q_{jt} + \gamma_{i} Q_{i,t-1} +\delta_{i} \sum_{j=1, j\neq i}^N w_{ij} Q_{j,t-1}+ \xV_{it}' \bV_{i}  ]$.
    $\Lambda$ is the matrix of $\lambda_{ij}$
     with the top $r\times r$ submatrix being lower triangular and positive diagonal elements.
    The state equation is $\fV_t=\Phi \fV_{t-1} + \uV_t$ with $u_t\sim N(0, I_t)$,
where $\Phi={\rm diag}\{\phi_1,...,\phi_r\}$.
    We apply the forward filtering and backward sampling method to draw from its posterior.

        \item A truncated normal can make a draw of
            $\phi_j\mid \cdot\sim N(\ol{m}_{\phi_j}, \ol{h}_{\phi_j}^{-1})\textbf{1}(|\phi_j|<1)$,
where $\ol{h}_{\phi_j}=\sum_{t=2}^{T}f^2_{j,t-1}$ and $\ol{m}_{\phi_j}=(\sum_{t=2}^{T}f_{j,t}f_{j,t-1})/(\sum_{t=2}^{T}f^2_{j,t-1})$ for $j=1,...,r$.
    \item $\Lambda$ is conditionally Gaussian with simple triangular
    identification restrictions.
    For $i\leq r$, denote it as $\betaB_i=(\lambda_{i1},...,\lambda_{ii})'$ with $\lambda_{ii}>0$; otherwise,
     we randomly draw a vector of length $r$ and denote it as $\betaB_i=(\lambda_{i1},...,\lambda_{ir})'$.
    The conditional distribution of $\betaB_i$ can be summarised in a linear regression as
    $y^*_i = X^*_i \betaB_i + \eV^q_i$,    where $y^*_i$ is the collection of    $y^*_{it}=Q_{it}- [\rho_{i} \sum_{j=1, j\neq i}^N w_{ij} Q_{jt} + \gamma_{i} Q_{i,t-1} +\delta_{i} \sum_{j=1, j\neq i}^N w_{ij} Q_{j,t-1}+ \xV_{it}' \bV_{i} ]$.
    If $i\leq r$, $X_i^*$ is the first $i$ columns from the matrix of factors $F$,
    otherwise, $X_i^*$ is simply the full $F$ matrix.
    We draw $\betaB_i$ from $N(\ol{\mV}_{\beta_i}, \ol{H}_{\beta_i}^{-1})$, where
    $\ol{H}_{\beta_i}=H_{\beta_i}+\sigma^{-2}_q(X^*_i)'X^*_i$ and $\ol{\mV}_{\beta_i}=\ol{H}_{\beta_i}^{-1}(\sigma^{-2}_qX^*_i)'\yV^{*}_i$.
    The $H_{\beta_i}$ is implied by the prior of $\Lambda$.

    \item We apply a Metropolis-Hastings algorithm to sample $\sigma$ based on
    $y_{it} = Q_{it}+\varepsilon_{it}$, where $\varepsilon_{it} \sim ALD(\tau, \sigma)$.
    \item We sample $V_{it}$ by $V_{it}=x^{-1}$ with $x \sim IG(\mu, \lambda)$,
        where $IG$ means inverse Gaussian with
$\mu = \sqrt{ \frac{(\xi_1^2 + 2\xi_2^2)\sigma^2}{(y^*)^2_{it}}}$,
$\lambda=\frac{\xi_1^2 + 2\xi_2^2}{\xi_2^2}$ with $\xi_1=\frac{1-2\tau}{\tau(1-\tau)}$, $\xi_2=\sqrt{\frac{2}{\tau(1-\tau)}}$.

      \item
      The $\QV_t$ has a state space representation.
    The state equation is
    $\QV_t = B\QV_{t-1} + X_t + \uV^q_t$ with $\uV^q_t\sim N(0, \Sigma)$,
    where  $A = (I-\rhoB W)^{-1}$, $B=(I-\rhoB W)^{-1} (\gammaB+\deltaB W)$,  $X_t=(I-\rhoB W)^{-1} \left([xb]_t + \Lambda \fV_t\right)$
 and $\Sigma=\sigma^2_q AA'$.
 The measurement equation is $\YV_t = \QV_t + \uV_t + \uV^y_t$,
 where $\uV_t$ is the vectorisation of $\xi_1u_{it}\equiv \xi_1\sigma V_{it}$ and $\uV^y_t\sim N(0, D_t)$ with $D_t=diag(\xi_2^2\sigma^2 V_{1t},...,\xi_2^2\sigma^2 V_{Nt})$.
    We apply the forward-filtering and back-ward sampling method.

    \end{enumerate}


\begin{proposition}
The proposed MCMC sampling procedure, using equation (\ref{eq:sq:q}), can generate posterior
samples from the distribution $\pi(\vartheta_\tau \mid X, Y)$ in (\ref{Posterior}) through an importance sampling procedure.
\end{proposition}

This section obtains the ``pseudo'' posterior $p(\vartheta_\tau | X, Y)$.
The exact posterior $\pi(\vartheta_\tau | X, Y)$ in (\ref{Posterior})
can be built through  importance sampling.  Our simulation study demonstrates that this computationally efficient MCMC approach yields satisfactory results.
For example, consider to compute the posterior expectation of $q(\vartheta_\tau)$ with $q(\cdot)$ being a known function of $\vartheta$:
\[
E\left[q(\vartheta_\tau)\mid X, Y\right]=
\int
q(\vartheta_\tau)
\pi(\vartheta_\tau\mid X, Y)d\vartheta_\tau.
\]
We can use the importance sampling:
\[
E\left[q(\vartheta_\tau)\mid X, Y\right]=
\frac{
\int
q(\vartheta_\tau) \frac{
\pi_K(\vartheta_\tau\mid X, Y)}{p_K(\vartheta_\tau\mid X, Y)}
p(\vartheta_\tau\mid X, Y)d\vartheta_\tau
}
{
\int
\frac{
\pi_K(\vartheta_\tau\mid X, Y)}{p_K(\vartheta_\tau\mid X, Y)}
p(\vartheta_\tau\mid X, Y)d\vartheta_\tau,
}
\]
where
$\pi_K(\vartheta_\tau\mid X, Y)$ and
$p_K(\vartheta_\tau\mid X, Y)$
are the kernels of
$\pi(\vartheta_\tau\mid X, Y)$
and
$p(\vartheta_\tau\mid X, Y)$, respectively.
Because we have a set of posterior samples from
$p(\vartheta_\tau\mid X, Y)$
and the kernels are easy to compute,
the posterior expectation of
$q(\vartheta_\tau)$
can be obtained without having $\pi(\vartheta_\tau\mid X, Y)$ directly.
Appendix
E.3 in supplementary document demonstrates Proposition 1 through a small Monte Carlo simulation.

Note that computing the posterior kernel \( p(\vartheta_\tau \mid X, Y) \) can be  computationally expensive.
However, after obtaining a sample using our method, it remains feasible, as the importance sampling process is parallelisable.
A key aspect of successful importance sampling is having a good proposal distribution,
 and our method plays a critical role in ensuring that the posterior distribution is precisely targeted.

{\remark
This section provides a sufficient approach, particularly when \( \sigma^2_q \) is tuned for robustness.
An alternative way to address the computational burden in the importance sampling is to let $\sigma^2_q \rightarrow 0$ as the number of MCMC sampling increases.
Then, we can have the posterior converging to the true posterior density \( \pi(\vartheta_\tau \mid X, Y) \).
As a result, after a certain number of MCMC iterations, the generated sample
from
\( p(\vartheta_\tau \mid X, Y) \)
is regarded as the sample from the true posterior density \( \pi(\vartheta_\tau \mid X, Y) \).
}

\subsection{Number of Factors}
In Bayesian theory,
the number of factors can be viewed as a random variable.
Its posterior can be inferred from by exploring the marginal likelihood of
models with different value of $r$.
However, because of the high computational cost,
we adopt the idea of the sparse mixture approach if \cite{malsiner2016model}
by setting a large but finite number of factors.
Each factor has a ``switch'' variables taking value of 0 and 1
indicating whether the corresponding factor is selected.
This construction follows the literature of stochastic search
variable selection dated back to \cite{mitchell1988bayesian}.

In particular, we revise \eqref{eq:sq:q} to have
\begin{align}
 Q_{it}& = \rho_{i, \tau} \sum\limits_{j=1, j\neq i}^N w_{ij} Q_{jt}
 + \gamma_{i, \tau} Q_{i,t-1} +
 \delta_{i, \tau} \sum\limits_{j=1, j\neq i}^N w_{ij} Q_{j,t-1}+
 \xV_{it}' \bV_{i, \tau} + (\sV_{\tau}\circ \fV_{t, \tau})'\lambdaB_{i, \tau}  + e^q_{it, \tau},\label{eq:sq1:q}
\end{align}
where $\sV_{\tau}$ is a $r_{\max} \times 1$ vector of $0$'s and $1$'s
and the symbol $\circ$ means the Hadamard product.
The maximum number of factors is $r_{\max}$.
Each element $s_{j, \tau}$ in vector $\sV_{\tau}$ controls for whether factor $f_{j,\tau}$ is selected.

Define the prior of $\sV_{\tau}=(s_{1,\tau},...,s_{r_{\max},\tau})$ as
\[P(s_{j,\tau}=1)=\pi,\quad P(s_{j,\tau}=0)=1-\pi\]
for $j=1,...,r_{\max}$.
The posterior distribution of the number of factors is
the distribution of the sum of $\sV_{\tau}$.

Because the only difference between \eqref{eq:sq:q} and \eqref{eq:sq1:q} is $\sV$,
we only need to add one more step for $\sV_{u_{it}}$ and revise the step of $F$ slightly
in the original MCMC algorithm to infer
the number of factors as follows.
\begin{enumerate}
    \item
    Draw $s_{k, \tau}$ from a simple Bernoulli distribution.
    Define
    \[y^*_{it} = Q_{it} -\left[ \rho_{i,\tau} \sum\limits_{j=1, j\neq i}^N w_{ij} Q_{jt}
 + \gamma_{i,\tau} Q_{i,t-1} +\delta_{i,\tau} \sum\limits_{j=1, j\neq i}^N w_{ij} Q_{j,t-1}+ \xV_{it}'\bV_{i,\tau} + (\sV_{\tau}\circ \fV_{t,\tau})'\lambdaB_{i,\tau} \right]\]
\[p(s_{k,\tau}=m\mid \cdot)\propto p(s_{k,\tau}=m) \prod\limits_{i=1}^N\prod\limits_{t=1}^T f_N(y^*_{it}\mid 0, \sigma^2_q),\]
where $m=0$ or $1$,
with the corresponding $s_{k,\tau}$ being set to $0$ or $1$.
There is a slight abuse of notation.
Namely, $y^*_{it}$ is different when $s_{k,\tau}=0$ and $1$.
\item To draw $\fV_{t,\tau}$,
we split this into two parts.
\begin{enumerate}
    \item Only draw the active factors for $s_{k,\tau}=1$.
    \item Conditional on the active factors,
    draw the other factors similar as in the MCMC step. This approach follows the Reversible jump MCMC method in \cite{green1995reversible}.
    \end{enumerate}
\end{enumerate}

{
\remark
In practice, we tune $\sigma^2_q$
to control an $R^2$ level from \eqref{eq:sq:q} or \eqref{eq:sq1:q}.
Alternatively, if we do not tune but want estimate $\sigma^2_q$
 we can monitor the $R^2$ instead.
For a high $R^2$ (in our application more than $98\%$),
the parameters can be used as if they were drawn from
the true posterior of the model.
Alternatively, one can use the posterior sample from this model setting
as a proposal distribution for an importance sampling scheme
applied to the original model.
Because the importance sampling is parallelisable, such second stage computation is much more affordable than any generic methods.
}

\section{Asymptotic results}

To provide a theoretical justification for the Bayesian method, this section presents results on posterior consistency. A sequence of posterior distributions is considered consistent if, as the length of the time series and the number of cross-sectional units increase, the posterior converges to the degenerate measure at the true parameter value of the population density. Intuitively, posterior consistency ensures that the information from the quantile objective function outweighs the prior information. Before analysing the asymptotic behavior of the posterior distribution, we first need to examine the average consistency of the frequentist estimator. This is necessary to establish a set of proper conditions (conditions that have not yet been fully explored) that guarantee the convergence of the estimated model to the true population density.

\subsection{Assumptions}

Below, we denote the true spatial parameters, the lag coefficients and the true regression coefficient as $\rho_{i,0,\tau}$, $\delta_{i,0,\tau}$, $\gamma_{i,0,\tau}$  and $\bV_{i,0,\tau}$, respectively.
Similarly, we denote $F_{0,\tau}=(\fV_{1,0,\tau},...,\fV_{T,0,\tau})'$ and $\Lambda_{0,\tau}=(\lambdaB_{1,0,\tau},...,\lambdaB_{N,0,\tau})'$ as the true factors and loadings. A set of regularity conditions that are needed for theoretical analysis are given as follows.

\subsubsection*{Assumption A: Common factors}

\noindent
Let $\mathcal{F}$ be a compact subset of $R^{r_\tau}$.
The common factors $\fV_{t,0,\tau}\in \mathcal{F}$
satisfy $T^{-1}\sum_{t=1}^T\fV_{t,0,\tau}\fV_{t,0,\tau}'=I_{r_\tau}$.

\subsubsection*{Assumption B: Factor loadings, the lag coefficients and regression coefficients}

\noindent
(B1)
Let $\mathcal P, \mathcal D, \mathcal G$ be compact subsets of $\mathbb R$, and let  $\mathcal{B}$ and $\mathcal{L}$ be compact subsets of $\mathbb R^{p+1}$ and $\mathbb R^{r_\tau}$, respectively.
The spatial parameters $\rho_{i,0,\tau}$ and $\delta_{i,0,\tau}$,
 the lag coefficients $\gamma_{i,0,\tau}$,
the regression coefficient $\bV_{i,0,\tau}$, and the factor-loading $\lambdaB_{i,0,\tau}$ satisfy that $\rho_{i,0,\tau}\in\mathcal P$,
$\delta_{i,0,\tau}\in\mathcal D$,
$\gamma_{i,0,\tau}\in\mathcal G$,
$\bV_{i,0,\tau}\in \mathcal{B}$ and $\lambdaB_{i,0,\tau}\in \mathcal{L}$ for each $i$.
\hangindent=30pt

\noindent
(B2)
The factor-loading matrix
$\Lambda_{0,\tau}= (\lambdaB_{1,0,\tau},\ldots,\lambdaB_{N,0,\tau})'$
satisfies $N^{-1}\sumiN\lambdaB_{i,0,\tau}\lambdaB_{i,0,\tau}'\xrightarrow{p} \Sigma_{\Lambda_\tau}$, where $ \Sigma_{\Lambda_\tau}$ is an $r_\tau\times r_\tau$ positive definite diagonal matrix with diagonal elements distinct and arranged in the descending order. In addition, the eigenvalues of $\Sigma_{\Lambda_\tau}$ are distinct.
\hangindent=30pt

\subsubsection*{Assumption C: Idiosyncratic error terms}

\noindent
(C1):
The random variable
\[\varepsilon_{it,\tau}=y_{it}-Q_{y_{it}}\Big(\tau| X_t, B_{\tau}, \fV_{t,\tau}, \Lambda_{\tau},\rhoB_{\tau}, \deltaB_{\tau}, \gammaB_{\tau}\Big)\]
satisfies $P(\varepsilon_{it,\tau}\le 0)=\tau$, and is independently distributed over $i$ and $t$,
conditional on $X_t$, $B_{0,\tau}$, $F_{0,\tau}$, $\Lambda_{0,\tau}$, $\rhoB_{0,\tau}$, $\deltaB_{0,\tau}$ and $\gammaB_{0,\tau}$.
\hangindent=30pt

\noindent
(C2): The conditional density function of $\varepsilon_{it,\tau}$ given $\{X_t, B_{0,\tau}, F_{0,\tau},\Lambda_{0,\tau}, \rhoB_{0,\tau}, \deltaB_{0,\tau}, \gammaB_{0,\tau}\}$,
denoted as $g_{it}(\varepsilon_{it,\tau})$, is continuous.
In addition, for any compact set $\mathcal{C}$, there exists a positive constant $\underline g>0$ (depending on $\mathcal{C}$)
such that $\inf_{c\in \mathcal{C}} g_{it}(c) \ge \underline g$ for all $i$ and $t$.
\hangindent=30pt

\subsubsection*{Assumption D: Weight matrix}

\hangindent=30pt
(D1): $W$ is an exogenous spatial weights matrix whose diagonal elements of $W$ are all zeros. In addition, $W$ is bounded by some constant $C$ for all $N$ under $\|\cdot\|_1$ and $\|\cdot\|_\infty$.
\hangindent=30pt

\noindent
(D2): The matrix $(I_N-\rhoB_\tau W)^{-1}$ satisfies $\sup_{\bm{\rho}_\tau \in\calP}\|I_N-\rhoB_\tau W\|_2 <C$, and
\[\sup_{\bm{\rho}_\tau \in\calP}\Big(\Big\|(I_N-\rhoB_\tau W)^{-1}\Big\|_1\vee\Big\|(I_N-\rhoB_\tau W)^{-1}\Big\|_\infty\Big)<C\]
 where $C$ is some positive constant.
\hangindent=30pt

\subsubsection*{Assumption E: Explanatory variables and design matrix}

\noindent
(E1):
For a positive constant $C$, explanatory variables satisfy $\sup_{it}\|\xV_{it}\|\le C$ almost surely.
\hangindent=30pt

\noindent
(E2): Denote $\bm{Q}_{t,\tau}^0=\bm{Q}_t(X_t, F_{t,0,\tau},B_{0,\tau},\Lambda_{0,\tau},\rhoB_{0,\tau},\gammaB_{0,\tau},\deltaB_{0,\tau})$. Let $\calX(B_{0,\tau})$ be an $N\times T$ matrix with its $(i,t)$-th entry $\xV_{it}'\bV_{0,\tau}$. Define $u_{it,0,\tau}$ to be the $(i,t)$th element of $\mathcal U_{0,\tau}$ with
\hangindent=30pt
 \[\mathcal U_{0,\tau}=W(I_N-\rhoB_{0,\tau}W)^{-1}
\Big[\calX(B_{0,\tau})+\Lambda_{0,\tau}F_{0,\tau}'+(\gammaB_{0,\tau}+\deltaB_{0,\tau}W)\mathcal Q_{-1,\tau}^0\Big].\]
where $\mathcal Q_{-1,\tau}^0=[\bmQ_{0,\tau}^0, \bmQ_{1,\tau}^0, \dots, \bmQ_{T-1,\tau}^0]$. Let $\zV_{it, \tau}=(u_{it,0,\tau}, Q_{i,t-1,\tau}^0, \sum_{j=1}^N w_{ij}Q_{j,t-1,\tau}^0, \xV_{it}')'$ with $Q_{i,t-1}^0$ being the $i$-th element of $\bm{Q}_{t-1}^0$, and $\ZV_{i,\tau}=(\zV_{i1,\tau}, \zV_{i2,\tau}, \dots, \zV_{iT,\tau})'$. Further define
$A_{i,\tau}=\frac{1}{T}\ZV_{i,\tau}'M_{F_\tau} \ZV_{i,\tau}$,
$B_{i,\tau}=(\lambdaB_{i,0,\tau}\lambdaB_{i,0,\tau}') \otimes I_T$,
$C_{i,\tau}=\frac{1}{\sqrt{T}}[\lambdaB_{i,0,\tau} \otimes (M_{F_\tau}\ZV_{i,\tau})]'$ with
$M_{F_\tau}=I-F_\tau(F_\tau'F_\tau)^{-1}F_\tau'$.
Let ${\mathcal{F}}_\tau$ be the collection of $F_\tau$ such that
${\mathcal{F}}_\tau=\{F_\tau:F_\tau'F_\tau/T=I_{r_\tau}\}$.
We assume that with probability approaching one,
\hangindent=30pt
\begin{eqnarray*}
\inf_{F_\tau\in {\mathcal{F}}_\tau} \lambda_{\min}\, \Big[ \frac1N \sum_{i=1}^N E_{i, \tau}(F_\tau) \Big]>0,
\end{eqnarray*}
where $\lambda_{\min}(A)$ denotes the smallest eigenvalue of matrix $A$, and $E_{i,\tau}(F_\tau)=B_{i,\tau}-C_{i,\tau}'A_{i,\tau}^{-1}C_{i,\tau}$.
\hangindent=30pt

\noindent
(E3): For each $i$, we assume that there exists a constant $c>0$ such that for each $i$, with probability approaching one,
\hangindent=30pt
\[\liminf_{T\to\infty}\lambda_{\min}\Big(\frac1T\ZV_{i,\tau}'M_{F_{0,\tau}}\ZV_{i,\tau}\Big)\ge c.\]

\subsubsection*{Assumption  F: Stationary condition}

The data generating process from (\ref{model}) is assumed to be stationary.
To ensure the stationarity, it is assumed that $\bar c <1$ where
\[\bar c =\sup_{\bm{\rho}_\tau\in\calP, \bm{\delta}_\tau\in\mathcal D, \bm{\gamma}_\tau\in\mathcal G}\Big|\lambda_{\max}\Big(A(\rhoB_\tau, \deltaB_\tau, \gammaB_\tau)\Big)\Big|<1.\]
where $\lambda_{\max}(A)$ denotes the eigenvalue of $A$ with the largest modulus.

{
\remark
Assumptions A and B on the factors and factor loadings are standard in factor models.
Similar to \cite{ando2020}, the factors and factor loadings are treated as parameters.
Assumptions C and D on the idiosyncratic errors and the spatial weighting matrix are standard assumptions in the literature.
Assumption E is necessary for deriving the consistency of the frequentist  estimator
(See \cite{ando2020} for similar assumptions).
Assumption F is a stationary condition similar to \cite{yu2008}.
Similar to the investigation in \cite{yu2008},
a sufficient condition for Assumption F is $\sup\|A(\rhoB_{\tau},\deltaB_{\tau},\gammaB_{\tau})\|_2 < 1$.
The stationary condition is verified accordingly in both simulation and empirical analysis.
}

Now, we investigate the consistency of the frequentist estimator, defined as the minimiser of (\ref{Allloss}) subject to a normalisation condition.
Recall $\vartheta_\tau=\{\rhoB_\tau, \deltaB_\tau, \gammaB_\tau, B_\tau, \Lambda_\tau, F_\tau\}$,
and $\hat{\vartheta}_\tau=\{\hat{\rhoB}_\tau, \hat{\deltaB}_\tau, \hat{\gammaB}_\tau, \hat{B}_\tau, \hat{\Lambda}_\tau, \hat{F}_\tau\}$ denotes the frequentist estimator.
A set of assumption A--F leads to the following result.

\begin{thm}
\label{thm1}
Suppose that the number of common factors in (\ref{model})
is correctly specified.
Under Assumptions A--F, $\log(N)/T\rightarrow 0$ as $N,T\rightarrow \infty$,
the frequestist estimator is the consistent estimator for their true values in the sense that
\begin{align}
&\frac1N\sum_{i=1}^N \|\hat{\rho}_{i,\tau}-\rho_{i,\tau,0}\|^2=O_p(\delta_{NT}^2),
&&\frac1N\sum_{i=1}^N \|\hat{\gamma}_{i,\tau}-\gamma_{i,\tau,0}\|^2=O_p(\delta_{NT}^2),
\nonumber\\
&\frac1N\sum_{i=1}^N \|\hat{\delta}_{i,\tau}-\delta_{i,\tau,0}\|^2=O_p(\delta_{NT}^2),
&&\frac1N\sum_{i=1}^N \|\hat{\bV}_{i,\tau}-\bV_{i,\tau,0}\|^2=O_p(\delta_{NT}^2),\nonumber\\
&\fracNT\|\hat\Lambda_\tau\hat F_\tau'-\Lambda_{\tau,0}F_{\tau,0}'\|^2=O_p(\delta_{NT}^2).\label{AverageRates}
\end{align}
where $\delta_{NT}=\max(\frac1{\sqrt N}, \frac1{\sqrt T})$.
\end{thm}

{\rm \remark
The last claim $\fracNT\|\hat\Lambda_\tau\hat F_\tau'-\Lambda_{\tau,0}F_{\tau,0}'\|^2=O_p(\delta_{NT}^2)$ further implies
\[\frac1N\sum_{i=1}^N \|\hat{\lambdaB}_{i,\tau}-\lambdaB_{i,\tau,0}\|^2=O_p(\delta_{NT}^2), \quad \frac1T\sum_{t=1}^T \|\hat{\fV}_{t,\tau}-\fV_{t,\tau,0}\|^2=O_p(\delta_{NT}^2). \]
}

{\rm \remark
The structure (\ref{modelreduced}) indicates that the quintile function at time $t$ is expressed as the sum of the previous quintile functions up to time $t-1$.
This created a difficult technical challenge to establish the claims in Theorem \ref{thm1}.
More specifically, as an intermediate result, the technical proof of our Lemma 2 (in the supplementary document) establishes
$\frac1N\sum_{i=1}^N \|\hat{\phi}_{i,\tau}-\phi_{i,\tau,0}\|^2=o_p(1)$,
where
$\phiB_{i,\tau,0}=(\rho_{i,\tau,0}, \gamma_{i,\tau,0}, \delta_{i,\tau,0}, \bV_{i,\tau,0}')'$,
$\hat\phiB_{i,\tau}=(\hat\rho_{i,\tau}, \hat\gamma_{i,\tau}, \hat\delta_{i,\tau}, \hat\bV_{i,\tau}')'$.
This result was not available from previous studies (\cite{ando2020}, \cite{chen2020}) due to the static nature of their model.
This illustrates one of the key challenges when we establish Theorem \ref{thm1}.

}

The next theorem also plays an important role when we investigate the consistency of our Bayesian MCMC procedure.
Theorem \ref{thmoverfit} implies that
it is ideal to set the number of common factors equal to or greater than the true number of common factors when one's focus is the consistent estimation of parameters $\rhoB_\tau$, $\deltaB_\tau$, $\gammaB_\tau$ and $B_\tau$.
To obtain the claim, we need additional assumption.

\subsubsection*{Assumption G: Identification of $B_\tau$ for over-fitted model}

\noindent
Let $F_\tau(k)$ be the common factor matrix with $k>r$ and $r$ being the true number of common factors,
$\mathcal Z_\tau(\phiB_\tau)$ be the $N\times T$ matrix with its $(i,t)$th entry equal to $\zV_{it,\tau}'\phiB_{i,\tau}$, where $\phiB_\tau=(\phiB_{1,\tau}, \phiB_{2,\tau}, \dots, \phiB_{N,\tau})'$ and $\phiB_{i,\tau}=(\rho_{i,\tau}, \gamma_{i,\tau}, \delta_{i,
\tau}, \bV_{i,\tau}')'$.
Here $\zV_{it,\tau}$ is defined in Assumption E.2.
There exists a positive constant $\breve c>0$ such that with probability approaching one,
\[
\inf_{F_{\tau}(k),F_{\tau}(k)'F_{\tau}(k)/T=I_k}
\frac1{NT}\|M_{\Lambda_{0,\tau}}\mathcal Z_{\tau}(\phiB_\tau)M_{F_{\tau}(k)}\|^2\ge \breve c \frac1N\sum_{i=1}^N\|\phiB_{i,\tau}\|^2,\]
where $M_{\Lambda_{0,\tau}}=I-\Lambda_{0,\tau}(\Lambda_{0,\tau}'\Lambda_{0,\tau})^{-1}\Lambda_{0,\tau}'$.

\begin{thm}
\label{thmoverfit}
Suppose that the specified number of common factors in (\ref{model})
is larger than the true number of common factors.
Under Assumptions A--G, $\log(N)/T\rightarrow 0$ as $N,T\rightarrow \infty$,
the corresponding frequestist estimator still satisfies the claims in (\ref{AverageRates}).
\end{thm}

Now, our concern is the sequence of posterior distributions
$\pi(\vartheta_\tau|Y,X)$ constructed by the size of $T\times N$ panel data,
generated from the true density $f(Y|X,\vartheta_{\tau,0})$.
In this paper, we show that the constructed posterior forms a
Hellinger-consistent sequence.
In regards to the posterior consistency based on Hellinger distance, we refer to
\cite{barron1999},
\cite{Ghosal},
\cite{walker2001}.

Recall the pseudo-likelihood based density function:
\[
f(Y|X,\vartheta_\tau)\propto
\exp
\left[ -\frac{1}{NT}
\sum_{i=1}^N\sum_{t=1}^T
q_{it,\tau}(\vartheta_\tau)
\right],
\]
where
$q_{it,\tau}(\vartheta_\tau)\equiv q_\tau(y_{it}-Q_{y_{it}}(\tau|X_t, F_{t,\tau},B_\tau,\Lambda_\tau,\rhoB_\tau,\gammaB_\tau,\deltaB_\tau))$.
We establish Bayesian consistency under the pseudo-likelihood $f(Y|X,\vartheta_\tau)$, in the sense that, for any $\mu>0$,
\begin{eqnarray}
\pi_{N,T,\tau}\left(\{\vartheta_\tau: H(f(Y|X,\vartheta_\tau),f(Y|X,\vartheta_{\tau,0}))>\mu\}\right) \rightarrow 0~~~{as}~N,T\rightarrow \infty,
\label{HC}
\end{eqnarray}
where $\vartheta_{\tau,0}$ is true value of $\vartheta_{\tau}$,
$\pi_{N,T,\tau}(\cdot)$ is defined as
\begin{eqnarray*}
\pi_{N,T,\tau}(A)=\int_A \frac{f(Y|X,\vartheta_\tau)}{f(Y|X,\vartheta_{\tau,0})}\pi(\vartheta_\tau)d\vartheta_\tau,
\end{eqnarray*}
where $A\subset \Theta$ is the subset of parameter space $\Theta$,
and
for two density functions $h(y)$ and $g(y)$, the Hellinger distance is defined as
$
H(h,g)=\left\{\int (g^{1/2}(y)-h^{1/2}(y)) d(y)\right\}^2
$.
To obtain the result in (\ref{HC}), we need an additional condition.

\subsubsection*{Assumption H: Kullback--Leibler property}

Let
$K_\varepsilon (\vartheta_{\tau,0})$ be
a Kullback--Leibler neighborhood of $\vartheta_{\tau,0}$ such that
$\vartheta_\tau$ satisfies
\begin{eqnarray*}
\int \log \{f(Y|X,\vartheta_{\tau,0})/f(Y|X,\vartheta_\tau)\} f(Y|X,\vartheta_{\tau,0})dY <\varepsilon.
\end{eqnarray*}
Then, the prior density $\pi(\vartheta_\tau)$ assigns  positive mass on all
Kullback--Leibler neighborhoods of the pseudo-likelihood based density under the true value $f(Y|X,\vartheta_{\tau,0})$,
$
\pi (K_v (\vartheta_{\tau,0}))>0$.

\begin{thm}
Under Assumptions A--H, as $N,T$ go to infinity with $N/T\rightarrow 0$,  then result (\ref{HC}) holds.
\end{thm}

When Bayesian consistency is considered important, the above theorem offers guidance for designing an appropriate prior distribution. The prior distribution discussed in Section 3 is specifically constructed to satisfy Assumption H. As a result, we expect that the posterior mean under our prior will converge to the true parameter values as both $N$ and $T$
tend to infinity. This expectation is confirmed through our simulation study.

{\remark
A common question is about the asymptotic properties of the posterior distribution regarding the number of common factors in our MCMC procedure. When the number of common factors is strictly smaller than the true number, there exists a positive constant such that the expected quantile loss is larger than that under the true number of common factors. As a result, our MCMC procedure asymptotically eliminates posterior samples with $r$ smaller than the true number of common factors as both $N$ and $T$ tends to infinity.
}


\section{Analysis of gasoline price}

We apply our method to the fuel prices reported by retailers in Queensland, a state located in the northeast of Australia. In Queensland, an aggregation system for fuel price reporting has been established under Section 4 of the Fair Trading (Fuel Price Reporting) Regulation 2018.\footnote{See https://www.epw.qld.gov.au/about/initiatives/fuel-price-reporting} As a result, all fuel retailers in Queensland (including all fuel stations) are required to report their fuel prices as part of the Queensland fuel price reporting scheme, which helps motorists find the cheapest fuel prices. This requirement has been in effect since 3 December 2018. The data is publicly available at \url{https://www.data.qld.gov.au/dataset/fuel-price-reporting}.

\begin{figure}[h]
    \centering
    \begin{minipage}[t]{0.48\textwidth}
        \centering
        \includegraphics[width=\linewidth]{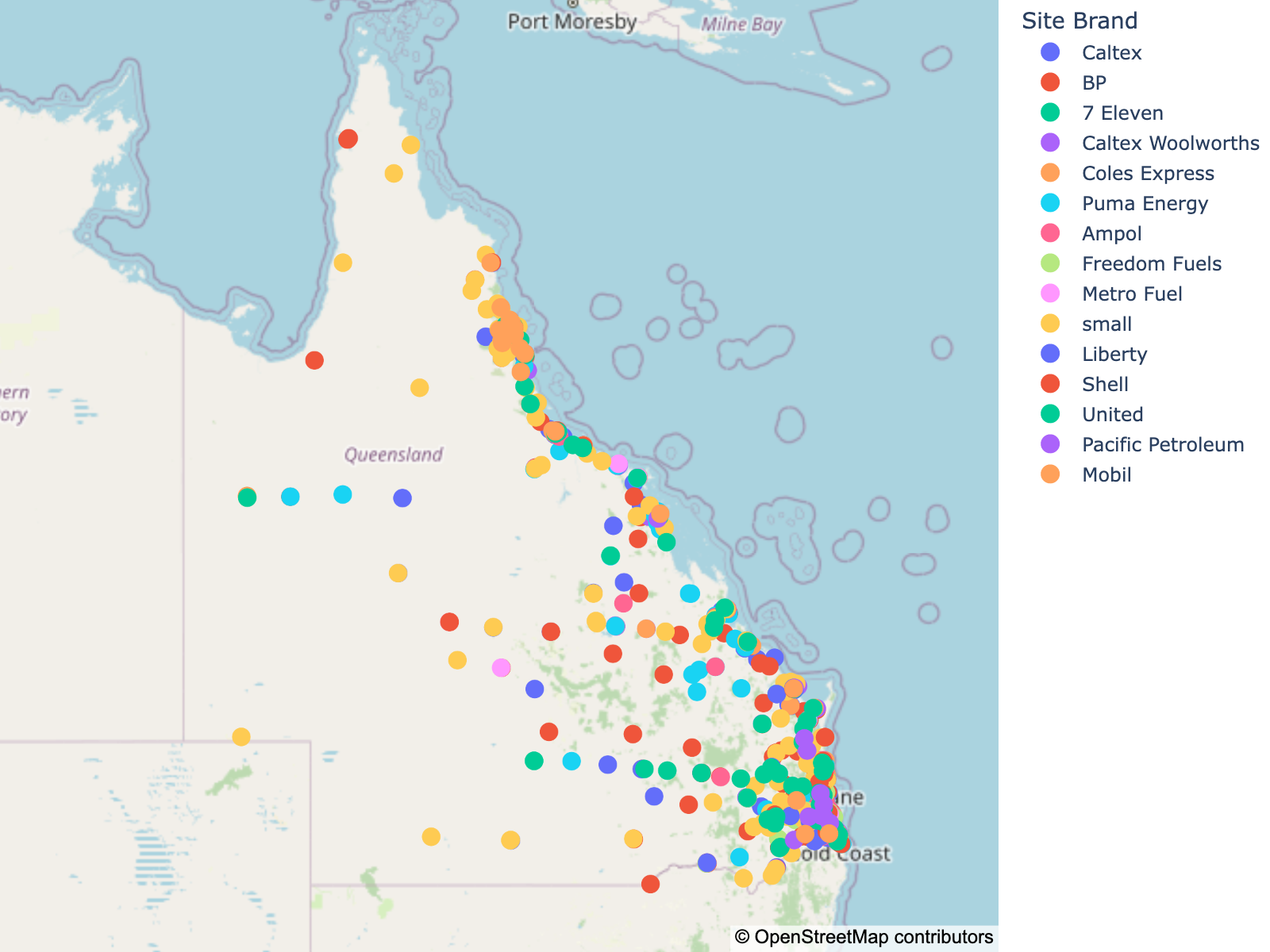}\\
        (a) Queensland Fuel Stations
    \end{minipage}
    \hfill
    \begin{minipage}[t]{0.48\textwidth}
        \centering
        \includegraphics[width=\linewidth]{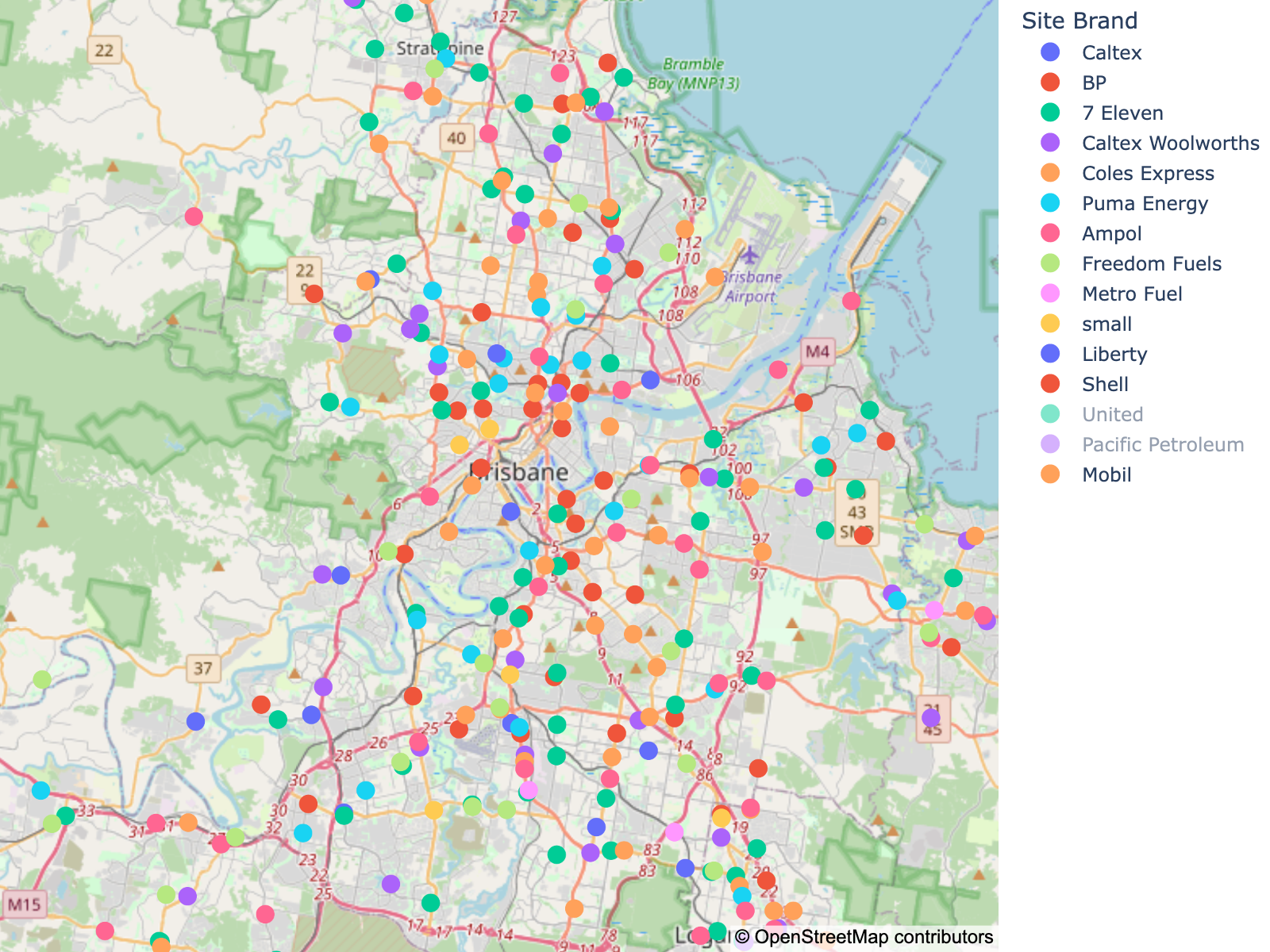}\\
        (b) Brisbane (Queensland's Capital City) Fuel Stations
    \end{minipage}
    \caption{
        Maps of fuel stations: (a) Queensland, (b) Brisbane.
    }
    \label{fig:stations}
\end{figure}

Figure~\ref{fig:stations} (a) shows the locations of these stations across Queensland, as well as their brands.\footnote{The brand \emph{small} aggregates the smaller brands with fewer than 5 stations.} In total, there are 1011 registered stations in Queensland.
Figure~\ref{fig:stations} (b) displays the locations of fuel stations in Brisbane, the capital city of Queensland, along with their respective brands. It is evident that many stations are situated close to one another, particularly in the municipal area.

\vspace*{3mm}

\subsection{Data}

We analyse unleaded gasoline, as in \cite{pinkse2002spatial}, because it has the largest market share and is nearly a homogeneous product.
 Our data spans from February 1, 2019, to September 30, 2021,
with a total of \( T = 973 \) days and no missing values. This sample includes five lockdown periods in Queensland
during the Covid-19 pandemic.

\begin{figure}[h]
    \centering
    \includegraphics[width=0.8\linewidth]{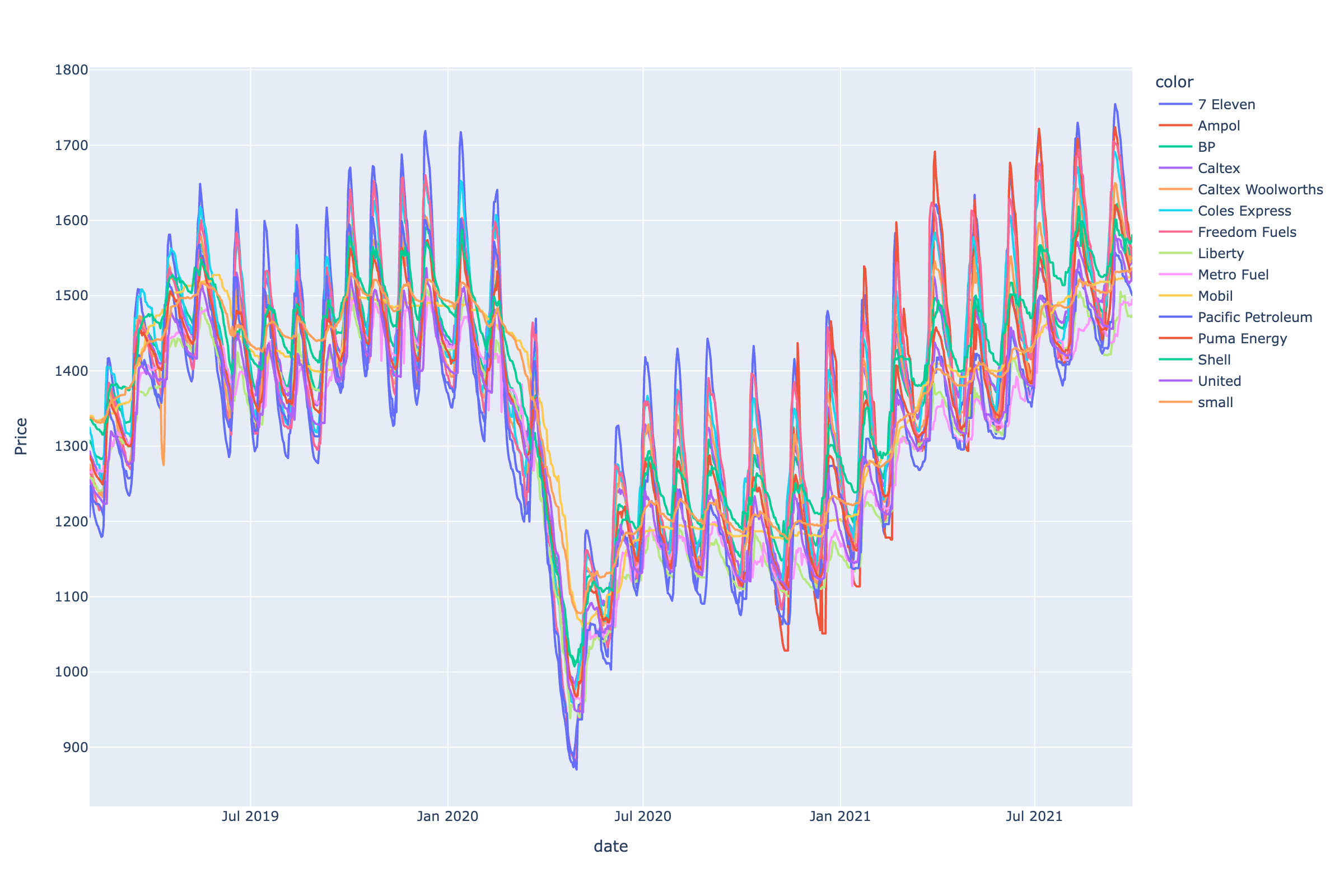}
    \caption{Average price by brands over time}
    \label{fig:price_time}
    The price unit is 1/10 cent per liter. Namely,
    1200 means 1.2 Australian dollar per litre.
\end{figure}

\begin{figure}
    \centering
    \includegraphics[width=0.5\linewidth]{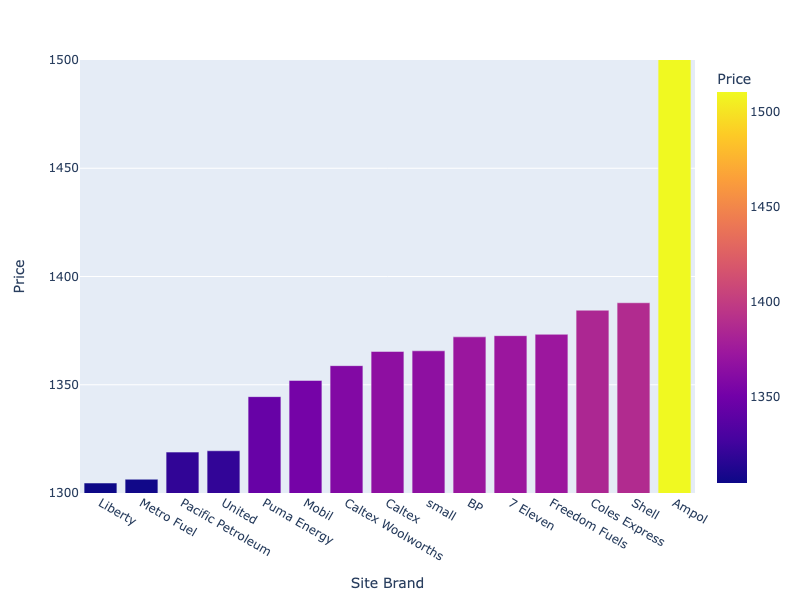}
    \caption{Average price by brands}
    \label{fig:price_brand}
\end{figure}

Figure~\ref{fig:price_time} shows the average fuel prices for each brand over time, revealing clear seasonality.
The significant drop in fuel prices in 2020 corresponds to the longest lockdown period, from March 26 to the end of April.
Additionally, when aggregating over time, Figure~\ref{fig:price_brand} highlights the price heterogeneity across brands.

In the following analysis, we exclude small brands with fewer than 5 stations and stations located on islands.
After this cleaning process, we are left with \( N = 946 \) stations. Throughout the sample period, no new stations
were built, nor were any stations decommissioned. While some stations changed brands, we account for the brand
effect in the estimation.

\subsection{Empirical model specification and estimation}

For the term  $\xV_{it}'\bV_{i,\tau}$ in (\ref{model}), we apply the following empirical specification:
\begin{align}
\xV_{it}'\bV_{i,\tau}
& =
{\rm Covid}_t \times b_{{\rm covid},i,\tau}+
{\rm Brand}_{it}\times b_{{\rm brand}_{it},\tau}
+
{\rm Year}_{t}\times b_{{\rm Year}_{it},\tau}+
{\rm Month}_{t}\times b_{{\rm Month}_{it},\tau}
\notag \\
&  \quad
+
{\rm Day~of~Week}_{t}\times b_{{\rm DayofWeek}_{it},\tau}
+G_{i,e_{it}}^{-1}(\tau)
\label{Empiricalmodel}
\end{align}

In the above equation, the subscript \(i\) refers to fuel station \(i\), and \(t\) represents time (day). The independent
variables in this model include several time effects: \({\rm Year}_t\), \({\rm Month}_t\), and \({\rm DayofWeek}_t\).
The year effect captures any overall trend, while the month and day-of-the-week effects capture explicit seasonal variations.
Each station \(i\) has its own parameter for these time variables. For instance, the effect of the year 2021 differs for station
1 compared to station 2.

The brand \({\rm Brand}_{it}\) dummy variables capture the brand effect. Note that some stations changed brands during the
sample period, which is why a double subscript is used for this variable. We include all brand dummy variables and exclude
the intercept for identification purposes, so the station-specific effect is naturally incorporated.

The variable \({\rm Covid}_t\) is a dummy variable that takes the value 1 if time \(t\) falls within a lockdown period
and 0 otherwise. Details of the lockdown periods are provided in the appendix. Briefly, there was only one long lockdown
period in 2020, from March 26 to the end of April, with all other periods being less than one week in duration.

For the weight matrix \( W \), we consider \textbf{driving} distance rather than geographic distance,
taking into account traffic conditions and speed limits in different areas.
To compute the average driving time between two stations, we use the Open Source Routing Machine (OSRM).
In this application, if two stations are within a 5-minute driving distance of each other, they are classified as neighbors.
We normalise each row of the matrix \( W \) such that the sum of the elements in each row equals 1.
Specifically, if station 1 has three neighbors (stations 2, 7, and 9), then
\( W_{1,2} = W_{1,7} = W_{1,9} = \frac{1}{3} \), and all other \( W_{1,j} = 0 \).

The prior is set to be informative but covers a broad range of the parameter space, as in the simulation.
The detailed settings can be found in the appendix.

\subsection{Results}

Due to the large number of parameters in the model, such as the number of \( \rho \), \( \delta \), \( \gamma \), and \( b \), \( \Lambda \), \( F \), which totals
\( 3N + N(p+1) + r_f(T + N) = 44,597 \) in our application, we do not store all simulated values during the MCMC process.
Instead, we focus on the posterior means, which allows us to accumulate values with minimal memory usage.
This approach makes it straightforward to evaluate uncertainties.
For example, if posterior variance is required, we can save the sum of the squared values,
then use the sample mean of the squared values and the sample mean to compute the sample variance.
Any moment-based posterior statistics can be derived in this way.

\begin{figure}[h]
    \centering
    \includegraphics[width=0.8\linewidth]{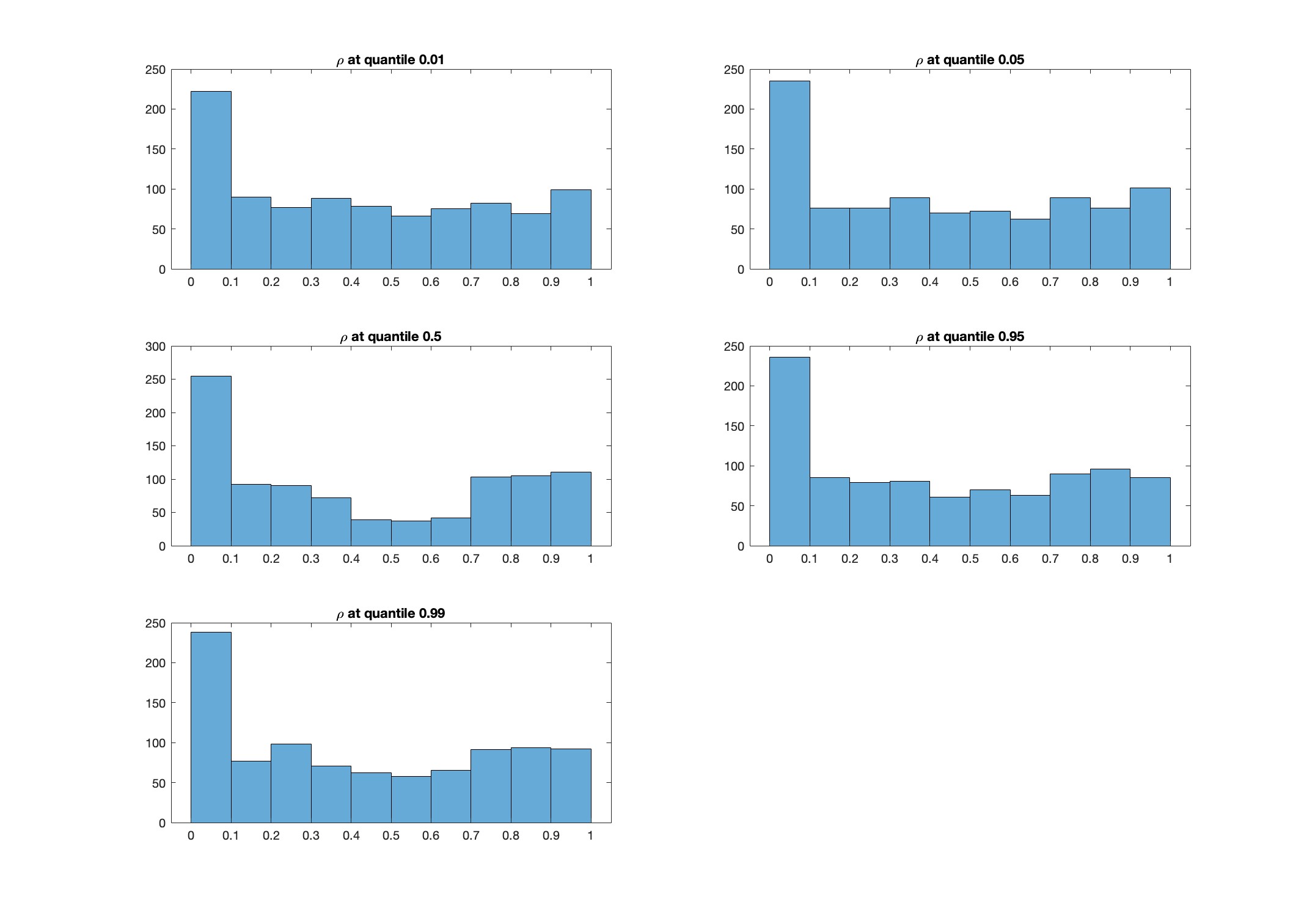}
    \caption{Distributions of $\rho$}
    \label{fig:post_rho}
\end{figure}

Figure~\ref{fig:post_rho} presents the histogram of the posterior means of \( \rho_{i,\tau} \) for all \( i \) at the quantiles \( \tau = (0.01, 0.05, 0.5, 0.95, 0.99) \).
Without imposing any restrictions, the distribution of \( \rho_{t,\tau} \) reveals two key characteristics.
First, the values are positive, indicating that positive spillover effects exist between fuel prices at nearby stations.
Second, the heterogeneity patterns are consistent across different quantiles.
Some stations exhibit greater sensitivity to their neighbors (larger \( \rho \) values), while others show minimal sensitivity, with \( \rho \) values close to zero.
Figure~\ref{fig:post_rho} underscores the need for a heterogeneous coefficient model.

\begin{figure}
    \centering  \includegraphics[width=1\linewidth]{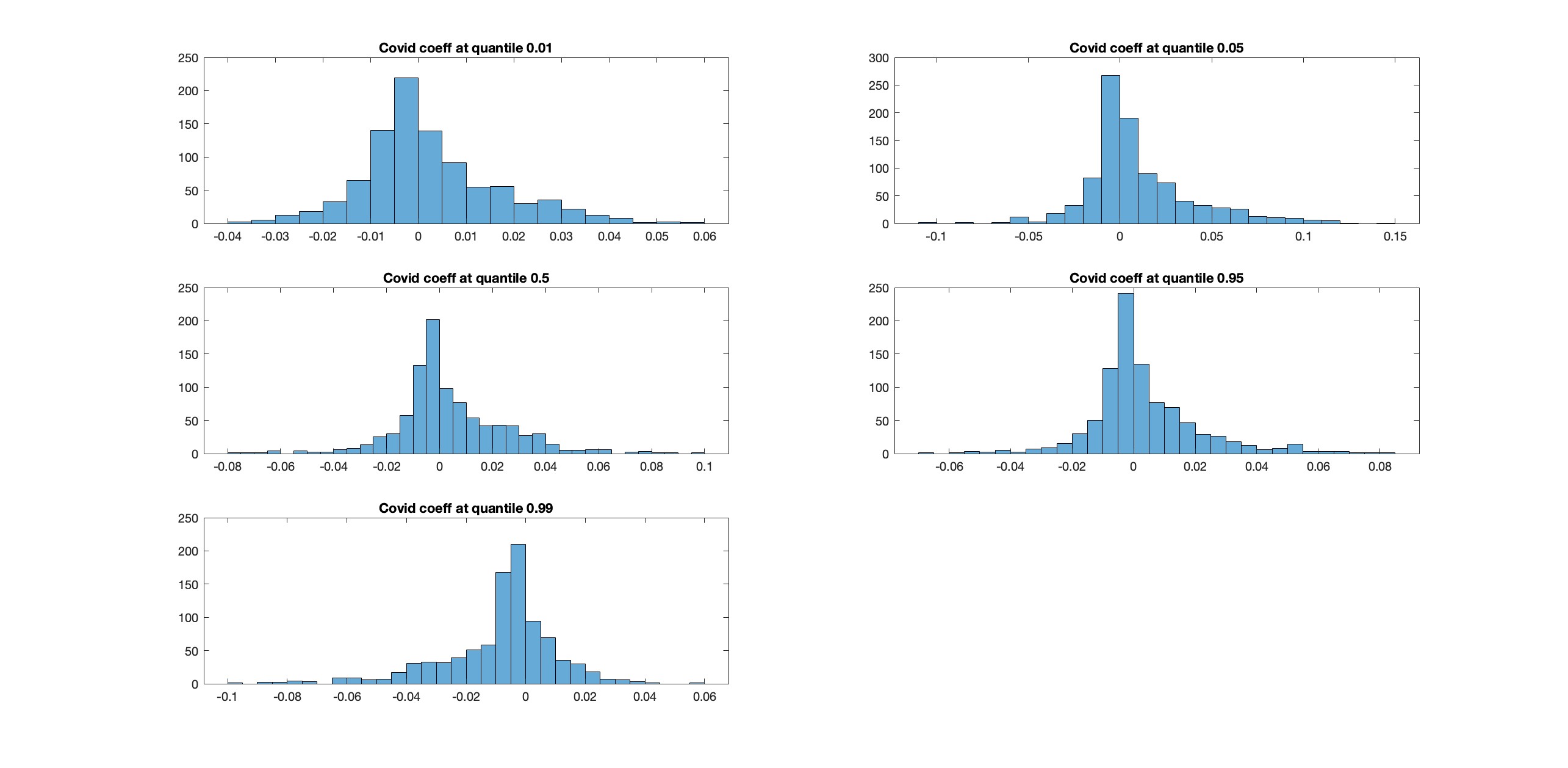}
    \caption{Distribution of Covid Coefficients}
    \label{fig:post_covid}
\end{figure}

Figure~\ref{fig:post_covid} shows the distribution of the posterior means
of the Covid lockdown coefficients at different quantiles. It is clear
that the distributions of these coefficients differ significantly across
quantiles. For instance, the histogram for \( \tau = 0.05 \) is
right-skewed, while the histogram for \( \tau = 0.99 \) is left-skewed.
The mode is around zero. This is not surprising, as the factors are intended
to capture any systematic price changes. Figure~\ref{fig:post_covid}
demonstrates that the lockdowns have caused changes in prices, not only
in terms of dispersion but also in how these price responses vary across
different quantiles.

\begin{figure}
    \centering  \includegraphics[width=1\linewidth]{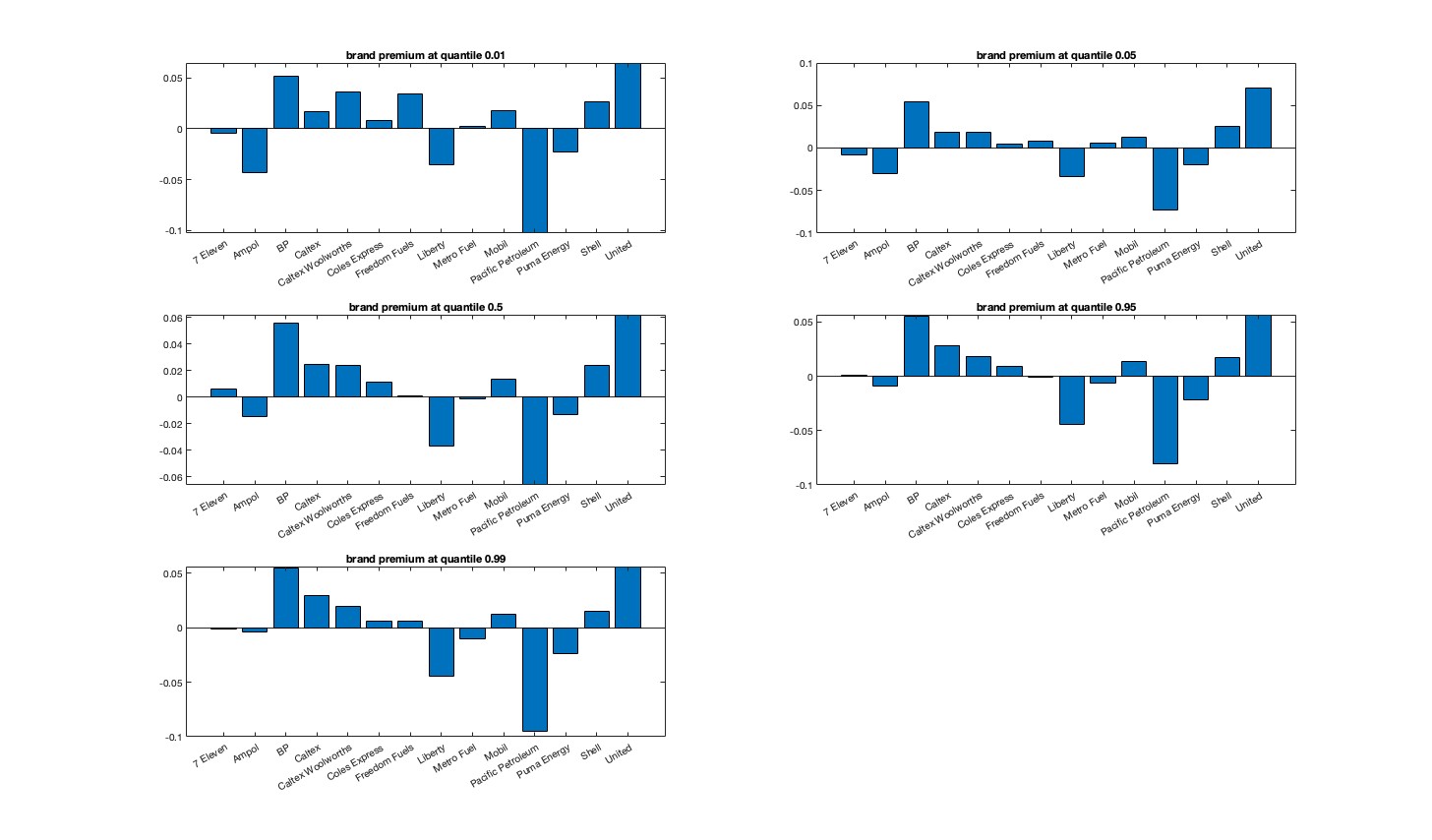}
    \caption{Average Brand Premiums}
    \label{fig:brand}
\end{figure}

Figure~\ref{fig:brand} displays the average brand premiums across
different quantiles. It is important to note that each station has its own
distinct brand premium, which can be interpreted as the average station
effect within the same brand. Similar patterns emerge across the brand
premiums. For instance, Pacific Petroleum consistently has the lowest
values, while United Petroleum consistently has the highest values
across all quantiles.

Similar to the spatial coefficient $\rho_{i,\tau}$, the lag coefficient
$\gamma_{i,\tau}$ and the lag-spatial coefficient $\delta_{i,\tau}$
demonstrate a strong pattern of heterogeneity, while exhibiting similar
patterns across quantiles, respectively. Due to page limitations, these
are not shown here.

\begin{figure}
    \centering  \includegraphics[width=1\linewidth]{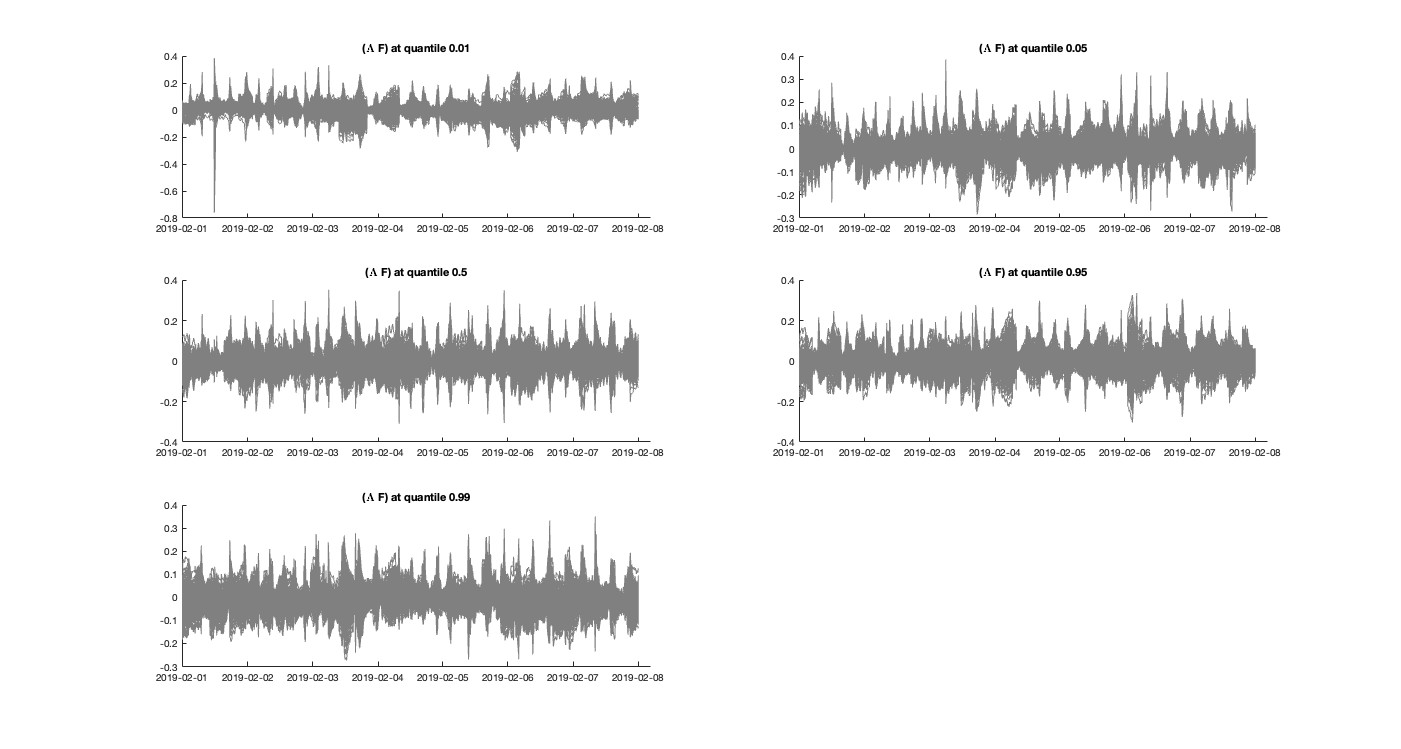}
    \caption{Posterior mean of $\Lambda F$}
    \label{fig:LF}
\end{figure}

The systemic factors and their effects on each time series are plotted
in Figure~\ref{fig:LF}. Each subplot represents $N=946$ time series,
depicting $\lambdaB_{i,\tau}'\fV_{t,\tau}$. It is evident that the factor
structure captures the seasonality present in the data. Since the simulation
study does not indicate the correct number of factors, we refrain from
analysing individual factors in this application.

We compute the average variation from the posterior mean of the contemporaneous effect. Specifically, we consider the term
$\rho_{i,\tau} \sum_{i \neq j, j = 1}^N w_{ij} Q_{y_{jt}} \left( \tau | X_t, F_{t,\tau}, B_\tau, \Lambda_\tau, \rhoB_\tau, \gammaB_\tau, \deltaB_\tau \right)$ in \eqref{model}.
The ratio of the average variation of this term to the average variation of the posterior values of the quantiles is approximately $30\%$ for all quantiles in $(0.01, 0.05, 0.5, 0.95, 0.99)$.
This suggests that the contemporaneous spatial effect plays a significant role in explaining the quantiles.

\section{Conclusion}

In this paper, we proposed a novel dynamic spatial panel quantile model with interactive effects. The model effectively captures several complex features simultaneously, including spatial spillover effects, heterogeneous regression coefficients, and unobserved heterogeneity that vary across quantiles. To estimate the model, we developed a Bayesian MCMC procedure capable of handling all these aspects, and we established Bayesian consistency to support its theoretical validity. The practical utility of the method was demonstrated through an application to gasoline price data in Australia.

\section*{Supplementary Materials}

All technical proofs of theoretical results and some numerical results are delegated to the supplementary document.

\baselineskip=21.3pt

\bibliography{ref2}

\end{document}